\documentclass[preprint]{vgtc}               

\graphicspath{{figures/}{pictures/}{images/}{./}} 

\usepackage{times}                     

\usepackage{tabu}                      
\usepackage{booktabs}                  
\usepackage{lipsum}                    
\usepackage{mwe}                       
\usepackage{cite}

\usepackage{mathptmx}                  

\onlineid{0}

\vgtccategory{Research}

\vgtcinsertpkg

\preprinttext{This is the authors’ preprint version of this paper. License: CC-By Attribution 4.0 International.}

\title{Inheriting the Count: How Visualization Literacy Got Its Measure}

\author{José Bener\thanks{e-mail: jose.bener@liu.se}\\ %
        \scriptsize Linköping University %
\and Miriah Meyer\thanks{e-mail: miriah.meyer@liu.se}\\ %
     \parbox{1.4in}{\scriptsize \centering Linköping University}}
     
\abstract{
Foundational frameworks in visualization have operationalized literacy as an individual competency, measured through chart-comprehension tasks. This focus raises a question: why has measurement become the dominant frame for understanding literacy? Rather than asking whether literacy should be measured, we ask how measurement became the field's way of understanding it. We trace visualization literacy back through the history of textual literacy and argue that in adopting the term, the field imported three values rooted in early government statistics: quantification, individualization, and binary classification. To mark the shift this history makes possible, we distinguish two waves: a first centered on assessment and individual proficiency, and a second that treats literacy as a situated practice, attentive to how people use visualizations in context and to the purposes they serve. Examining the construct this way suggests that refining assessment instruments does not, by itself, settle what literacy is, and points to directions for second-wave research: formative studies of literacy in context, culturally grounded instruments, and critical reading. This paper shows that visualization literacy inherited its measurement frame rather than discovering it, offers a two-wave vocabulary distinguishing assessment-based literacy from situated practice, and connects the field to a critical-literacy tradition with concrete alternatives to measurement. These contributions allow us to treat measurement and meaning as two parts of a single question, shifting attention from who counts as literate to what literacy is meant to do.

} 


\teaser{
  \centering
  \makebox[\textwidth][c]{%
    \includegraphics[
      width=1\textwidth,
      alt={A sequential overview of visualization literacy's inherited assessment values.}
    ]{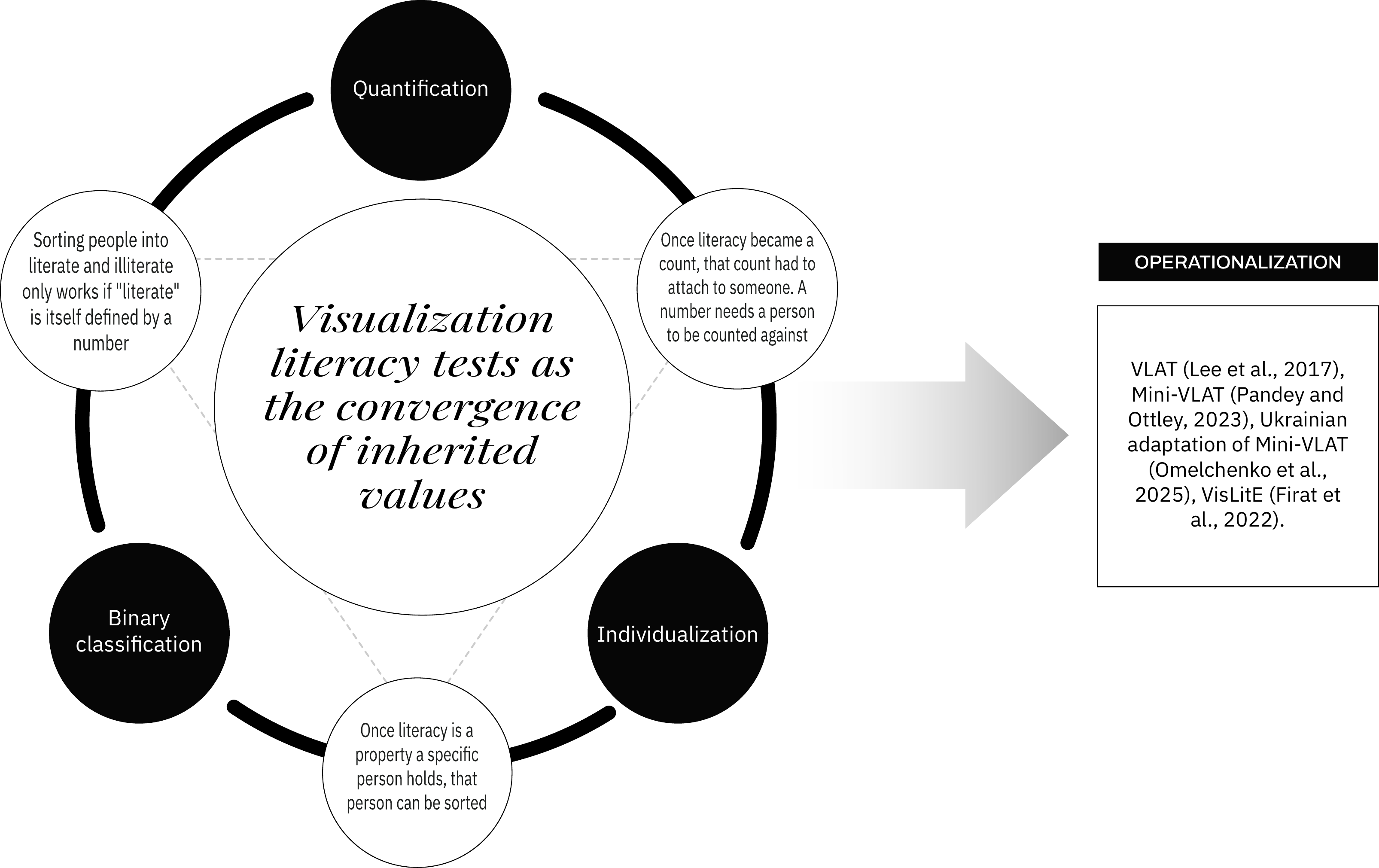}%
  }
  \caption{%
    Visualization literacy tests as the convergence of three inherited values. Quantification, individualization, and binary classification are shown here as interconnected, each positioned in relation to the others. This convergence is operationalized in assessment instruments including VLAT.%
  }
  \label{fig:teaser}
}

\begin{document}


\firstsection{Introduction}

\maketitle

Most research on visualization literacy focuses on an individual's ability to interpret graphical data representations, typically evaluated through assessments that identify values, compare quantities, and extract trends \cite{Boy:2014:PWA,Varona:2025:VL}. Recent critiques have begun to question this assessment-centered framing. Ge et al. \cite{Ge:2026:AVL}, for instance, call visualization literacy measurement a \textit{wicked problem}, too multifaceted to capture cleanly, and suggest that the search for a universal, generalizable test may itself be influencing how the community understands the construct.

Like these critiques, we see assessment tools as doing more than evaluating visualization literacy; they help define what the field understands literacy to be. Existing work has begun to examine this effect, asking how well assessments capture what we mean by literacy \cite{Hedayati:2024:FPP,nobre} and, most recently, whether literacy should be measured at all \cite{Solen:2025:VLS,akbaba2026value}. We set aside the question of whether measurement is effective or appropriate. Instead, we ask where the measurement frame came from: an inherited social project, not a natural analytical step. This raises two questions the field rarely asks: why has assessment become the dominant frame for visualization literacy, and what values do we accept when measurability defines what counts as literate?

To answer them, we trace visualization literacy back through the history of textual literacy. We draw on the nineteenth-century British case, among the first to make literacy countable, where political and administrative projects reduced situated practices of reading and writing to categories, counts, and comparisons \cite{Hacking:1990:TC,Vincent:2019:MHL}. To count literacy across a population, the state treated a varied practice as a single thing a person either had or lacked \cite{Vincent:2019:MHL}. Built on that simplification, the construct may not capture how people read and write, which could be why researchers have not agreed on a single literacy measure. Visualization inherited this construct \cite{Solen:2025:VLS}, yet still pursues a universal test \cite{Lee:2017:VLA}, so a similar limit may apply. We further argue that in adopting the term literacy, the field imported three values rooted in this history of government record-keeping: \textit{quantification}, turning literacy into a count; \textit{individualization}, tying that count to specific people; and \textit{binary classification}, sorting people into literate and illiterate.

This paper offers three contributions. First, we provide a historical account of how visualization literacy came to be defined and measured, tracing it through the history of textual literacy. Making this lineage explicit lets researchers ask whether visualization's measurement assumptions hold across cultural and institutional contexts. Second, we introduce a vocabulary of \textit{two waves} of visualization literacy. Other fields have used paradigm- and wave-framings to mark shifts in emphasis \cite{harrison2007three}; we use \textit{waves} in this lighter sense, to distinguish a first wave focused on assessment and individual proficiency from a second that treats literacy as situated practice. Third, we connect this work to the critical-literacy tradition, drawing on its long scholarship on alternatives to measurement to suggest concrete starting points for visualization research: formative assessment, situated materials, and critical reading practices.

By tracing the origins of literacy measurement, instead of only refining the instruments used to measure it, the visualization field can focus on questions that fall outside what the assessment approach can easily measure: how people use visualizations in context, whose purposes a visualization serves, and what literacy in visualization is ultimately meant to do.

\section{Related Work}
In recent years, visualization researchers have increasingly turned to history as a lens on the field, tracing familiar terms and ideals back to the values and contexts that produced them, and showing them to be situated choices, not neutral conventions. Akbaba et al. \cite{akbaba2021manifesto} trace \textit{chartjunk} to Tufte's minimalist values, noting that the community adopted the term without attending to the values built into it. Klein \cite{Klein:2024:VAP,klein2022what} connects the ideals of clarity and efficiency to William Playfair's political aims and recovers Elizabeth Palmer Peabody's charts, a tradition that later visualization overlooked. D'Ignazio and Klein's \textit{Data Feminism} \cite{DIgnazio:2020:DF} asks whose perspectives data practices recognize and highlights practitioners overlooked by standard histories, while work on data physicalization has revived pre-digital traditions of making with data \cite{Huron:2022:MWD}. Saharan et al. \cite{valueDATA} trace the values of universality, objectivity, and efficiency from the field's early figures into current practice and call for a more pluralistic approach. We pursue a similar strategy for the construct of literacy. Rather than treating literacy's values as inherent to visualization, we trace how they developed through the history of the word itself.

Among the works that reconsider the framing of visualization literacy, Akbaba's provocation \cite{akbaba2026value} sits closest to our position, questioning whether literacy should be measured at all and situating that question within debates about who defines literacy, who benefits, and what role education plays. Akbaba works through feminist and entanglement epistemology; we work \textit{genealogically}, tracing how measurement became embedded in the concept of literacy over time. The two papers diverge on method: Akbaba interrogates the value of literacy measurement directly, while we trace the history behind its apparent inevitability.

In addition to these historical works, our study draws on two further literatures used in later sections. First, recent work on visualization literacy assessment \cite{CulturalMini,nobre,Solen:2025:VLS,Boy:2014:PWA,Lee:2017:VLA} informs our account of the first wave (Section 4). Second, scholarship on textual literacy \cite{Street:1995:SLL,Graff:1991:LM,Freire:2018:PO,Cazden:1996:PML} provides the historical genealogy (Section 3) and the alternatives to measurement (Section 5).

\section{What Visualization Literacy Inherits}

The three values named in the introduction (\textit{quantification}, \textit{individualization}, and \textit{binary classification}) do not originate in visualization; the genealogy below traces each to a specific history that precedes the formalization of our field. Treating these values as an inheritance, not a built-in feature of visualization, lets us ask what contrasting or complementary framings of literacy are possible. We proceed in three steps: we show how literacy became measurable, how textual-literacy scholarship developed alternatives that expand that framing, and how visualization research adopted the term and its measurement frame.

Following Akbaba et al. \cite{akbaba2025entanglements}, who distinguish genealogy from history, we offer the following as a genealogy: one of many possible accounts that thread together the ideas, instruments, and moments shaping the present literacy construct, rather than a single definitive history of it. The British case we discuss in this study is one instance of a broader pattern, not the only one.

\subsection{How literacy became measurable}

As a formal concept, literacy is relatively new in the Western world, emerging alongside the formalization of statistics and modern state record-keeping \cite{Vincent:2019:MHL,Cressy:1977:LOI,Galenson:1981:LAP}. Consider the British case in 1836, when a government act mandated that every couple register their marriage. Each individual was required either to sign the register, if literate, or to make a mark if unable to write. This ordinary administrative act quietly transformed a marriage record into a measurement tool for literacy. What had served as documentation of civil union became an indicator of writing ability within the population.

In 1840, a half-page table published by the General Register Office of England and Wales changed how reading and writing were understood. Its columns reported the proportion of persons married in 1839 who had signed with marks. By counting the register entries this way, the Office's staff realized that summing the entries could \textit{``represent on one sheet of paper the communication capacity of an entire society''}~\cite{Vincent:2019:MHL}. A signature became a proxy for the ability to read and write, and an administrative record, now population-level data, began to provide, as officials put it, \textit{``much light upon the state of education with respect to writing among the adult population''}~\cite {Vincent:2019:MHL}.

We read this as the moment \textbf{quantification} emerged as a value, when the social phenomenon of literacy became countable. This shift is part of what Hacking calls the nineteenth-century \textit{avalanche of printed numbers,} as nation-states began to classify, count, and tabulate their subjects in new ways \cite{Hacking:1990:TC}. Literacy tables worked this way: by turning the decision to marry into a stable time series, the state could track and evaluate change, and these tables became, in Vincent's words, among \textit{``the first key performance indicators of public investment''} \cite{Vincent:2019:MHL}.

Once literacy became measurable, it could justify education budgets, judge reforms, and rank nations. This process, we argue, depends on flattening a complex human capacity into a single comparable number. As Hacking notes, counting does more than describe society; it creates the categories through which a state comes to know and govern its people \cite{Hacking:1990:TC}. This exchange, we argue, trades comparability for nuance, since a single number portrays literacy as more stable, bounded, and complete than it is, overlooking the partial, shared, and context-dependent ways people read, write, and communicate.

Alongside quantification, a second value begins to emerge: \textbf{individualization}. As Vincent puts it, \textit{``as it was counted, literacy was seen as a form of property, owned or absent, rather than one amongst a bundle of skills for living [...]''} \cite{Vincent:2019:MHL}. Literacy became an attribute assigned to a person, who could then be held responsible for having or lacking it, displacing the view of literacy as a social project that public policy makes more or less available.

For Foucault, this attachment of measures to individuals is the consequential move. The \textit{drawing up of tables} that organized economic flows, military registers, hospitals, and populations made it possible to supervise, assess, and rank the people recorded in them, the logic Foucault calls \textit{normalization} \cite{Foucault:1977:DP}. Once literacy is treated as an individual's recorded property, institutions can sort people by it: set a standard, rank everyone against it, and treat those below the line as the problem to be corrected. This, we argue, is why individualization matters for literacy: it relocates responsibility from social and material conditions to the person being measured, so that any shortcoming is attributed to the learner, not to the schools, materials, languages, communities, or systems around them.

We trace a third value, \textbf{binary classification}, to the record's form itself. A choice between signature and mark leaves no middle ground, and this binary, entangled with individualization, names the deficit before naming the capacity. The term illiteracy was in use decades before literacy entered common vernacular: \textit{``the inability to write appeared an absolute condition, and soon became described as `illiteracy'... `literacy' itself did not become a widely used term for another four decades''} \cite{Vincent:2019:MHL}.

The consequences of this value are difficult to set aside. Binary classification turns something complex and situated into a threshold, making literacy appear to be something a person simply has or lacks. In practice, people perform literacy differently across languages, contexts, technologies, and communities. In doing so, it makes partial, collaborative, critical, and context-specific forms of literacy harder to recognize.

Our genealogical reading of the British case lets us name the values we see entangled with the construct of literacy. In this reading, quantification became a value when signatures and marks made literacy countable; individualization, when the count attached to a person who could be ranked against others and held responsible for the result; and binary classification, when people could be sorted into those who possessed literacy and those who did not. These values are not inherently harmful; we problematize them because they narrow the meaning of literacy. They present it as an individual, countable, classifiable possession, not a situated practice shaped by social relations, material conditions, cultural contexts, and institutional responsibilities. Visualization inherited these values well before adopting the term, arriving with the word itself, and naming them as inherited, not intrinsic, is the first move toward asking whether they are values we want.

\subsection{The critical turn in textual literacy}

The measurement frame established in the nineteenth century carried into the institutional definitions of the twentieth, where literacy continued to be treated as a measurable technical skill. UNESCO's 1958 definition, for example, described a literate person as someone who could read and write a short, simple statement about everyday life \cite{UNESCO:2006:LFL}, treating literacy as a basic, observable competence. Street later names this skills-based framing the \textit{autonomous model of literacy}, in which literacy is a neutral, technical ability, not a socially situated practice \cite{Street:1995:SLL}.

Critical literacy scholars also challenged this technical model, most notably Freire~\cite{Freire:2018:PO}, who reframed literacy as a political, cultural, and world-making practice. Since the 1960s, much of this scholarship has questioned measurement-driven approaches, raising concerns about the literate/illiterate binary, the emphasis on individual skill, and the institutional uses of literacy assessment. Street's ideological model departs from the neutral, technical-skill view. It treats literacy as \textit{``an ideological practice, implicated in power relations and embedded in specific cultural meanings and practices''} \cite{Street:1995:SLL}. Freire and Macedo \cite{Freire:1987:RWW} likewise treat literacy as \textit{``a set of practices that functions to either empower or disempower people.''} Multiliteracies scholarship has produced curriculum models that ground literacy in learners' cultural and political environments instead of standardized items \cite{Cazden:1996:PML}. The same tradition also produced alternatives to measurement: formative assessment, which treats evaluation as ongoing feedback, not a one-off ranking \cite{Black:2009:DTA}, and New Literacy Studies, which views literacy as a social practice shaped by the habits of homes, schools, and communities \cite{Heath:2013:WW}. Visualization has drawn on little of this, borrowing the term \textit{literacy} and its values \cite{akbaba2026value} but not the critique that came with them. We develop these alternatives in Section~5; first, we turn to how visualization inherited the measurement frame.

\subsection{Inheriting measurement in visualization research}

As textual-literacy scholarship began to question its reliance on measurement, visualization researchers began to define what it means to be literate in our field. Concerns with graph comprehension and visual interpretation predate the term, but Boy et al. \cite{Boy:2014:PWA} were among the first to formalize visualization literacy as a distinct construct, defining it as \textit{``the ability to confidently use a given data visualization to translate questions specified in the data domain into visual queries in the visual domain, as well as interpreting visual patterns in the visual domain as properties in the data domain,''} a definition organized, like its textual inspiration, around acts of reading.

If Boy et al. provided a working definition of visualization literacy, then Lee, Kim, and Kwon \cite{Lee:2017:VLA} offered a way to operationalize it through the Visualization Literacy Assessment Test (VLAT). Their approach is grounded in a clear premise: VLAT rests on \textit{``a psychological assumption that human traits and skills can be quantified and measured.''} In building their assessment instrument, they narrow visualization literacy to the abilities their items can measure, which is the position Biesta \cite{biesta2009good} warns about: measuring what we value gives way to valuing what we can measure.

Seen this way, VLAT is a clear case of how the three inherited values of literacy settled into visualization research. Read against quantification, individualization, and binary classification, we argue that the test reproduces all three: it reduces a reader's ability to a single corrected score (\textit{quantification}); it tests one reader at a time and reports a percentile rank against everyone else (\textit{individualization}); and it establishes a \textit{binary classification}, since a threshold sorts readers into literate and illiterate.

Those who built the measure understood its cost; that understanding is what makes this inheritance striking. Victorian statisticians knew a signature was a crude proxy for literacy but counted anyway, judging the gain in comparability worth the loss \cite{Vincent:2019:MHL}. Lee et al. make the same admission that visualization literacy is \textit{``too complex to assign a single number,''} yet proceed because, in their view, a validated instrument is a necessary first step \cite{Lee:2017:VLA}. On our reading, then, the field did not merely inherit the values of textual literacy. It repeated, a century and a half later, the same knowing exchange of meaning for measurability.

That exchange has shaped the work that followed, and much of it has remained within this framework, extending it through assessment tools, models, and taxonomies \cite{Varona:2025:VL,Firat:2022:VLE,rodrigues}. In doing so, we suggest, visualization research reproduces a conception of literacy that Freire and Macedo criticized as \textit{``a mechanical process which overemphasizes the technical acquisition of reading and writing skills''} \cite{Freire:1987:RWW}, and inherits the questions literacy scholars have debated for decades: who defines literacy, how it should be measured, and whether it should be measured at all \cite{Freire:2018:PO}. If measurability comes from the history of textual literacy and not from literacy itself, refining an instrument is not, by itself, progress on what literacy is. A construct designed for population record-keeping \cite{Vincent:2019:MHL} strains to capture how people use visualization across contexts. Our notion of literacy is the product of a particular history \cite{Graff:1991:LM,Street:1995:SLL}, one that the field can contest in order to broaden it. The next section offers a vocabulary for doing so.

\section{Two Waves of Visualization Literacy}

We conceptualize literacy as entering visualization research in two distinct waves. We use the term \textit{waves} to indicate a shift in emphasis; we do not use it to designate an established body of work. The first wave characterizes the field's focus on measurability up to now, while the second wave identifies an emerging direction highlighted by recent critiques. Paradigm- and wave-framings circulate in visualization discourse: HCI has described its own evolution through shifting epistemologies~\cite{harrison2007three}, and visualization research has already engaged with paradigm and wave framing \cite{akbaba2025entanglements,frauenberger2019entanglement}. In our work, we do not claim a paradigm shift in the epistemology of visualization; we trace how the field has operationalized the single construct of literacy, and we suggest that its emphasis is shifting from measurement toward situated practice.

The \textbf{first wave} is organized around measurability and assessment. As Solen et al. \cite{Solen:2025:VLS} note, \textit{``a significant portion of visualization literacy work focuses on assessments, both their creation and application,''} typically testing whether readers can extract single values, compare values, or identify trends. VLAT is the canonical instrument of this wave \cite{Lee:2017:VLA}. Item response theory models follow the same logic \cite{Boy:2014:PWA}, as do recent taxonomies that map and refine the competency construct \cite{Varona:2025:VL,Firat:2022:VLE,rodrigues}. Their shared orientation treats literacy as something a person has more or less of, with the research task framed as measuring it well.

This orientation seems to embed values that the field has not explicitly chosen. It treats literacy as an individual property, tests it with standardized stimuli, and scores it quantitatively. Collective practice, situated stimuli, and qualitative accounts of reading fall outside this frame. The cost is not obvious from inside the frame, because a test taker who completes every VLAT task looks fully literate. Yet that same individual may be unable to ask why a chart was made, what it omits, or whose interests it serves, and we take these to be the questions that protect against being misled. A visualization can mislead precisely by being easy to decode \cite{pandey}, so fluency at extracting values offers little protection against a chart built to persuade. Solen et al. \cite{Solen:2025:VLS} name this gap as the distinction between visualization as communication and visualization as inquiry: the first wave attends to communication and has paid far less attention to inquiry. The literacy analogies, as they note, \textit{``imply a dichotomy, where one either has the ability or not,''} which is where the binary enters, even when the instrument that produced the score is continuous.

We propose that visualization literacy is taking a turn. The impulse to measure, formed in the context traced above, may be reaching its limits as new forms of engagement with data emerge. Treating literacy as something a person has or lacks arguably no longer offers a sufficient lens, and we see a second wave taking form. Part of this turn is already visible in the work that questions whether tests capture what matters \cite{Hedayati:2024:FPP,nobre,Solen:2025:VLS,akbaba2026value}. We propose that the turn be made explicit: a different relationship to literacy, not a refinement of measurement.

Drawing on critical pedagogy, the \textbf{second wave} would treat literacy as \textit{``the relationship of learners to the world''}~\cite{Freire:1987:RWW}. Visualizations are shaped by the contexts and interests that produce them~\cite{DIgnazio:2020:DF,Kennedy:2016:PPV}, which is why Freire and Macedo \cite{Freire:1987:RWW} treat literacy as \textit{``a form of cultural politics.''} Their conventions do rhetorical work, lending charts an air of objectivity that can mask the choices behind them~\cite{hullman,Kennedy:2016:PPV}. On this reading, the second wave begins when expert and non-expert users recognize a visualization as a social artifact \cite{politics}. This shifts literacy from individual competence toward situated and critical engagement. As the New London Group~\cite{Cazden:1996:PML} argues, \textit{``learning for this reason is also very social, as we rely on the artifacts of collective memory, and work with others in the essentially collaborative task of knowledge making.''}

In practice, this means paying attention to how visualizations persuade and whose interests they serve. A visualization can manipulate or misinform \cite{pandey}, serving interests that shape how data is read. The traditions that treat literacy as social practice and situated meaning-making \cite{Street:1995:SLL,Freire:1987:RWW,Cazden:1996:PML} recognize the kind of knowledge standardized instruments miss: reading a chart for what it shows and for whose purposes it serves. We suggest that visualization research can draw on this tradition to expand literacy beyond what is standardized and measurable.

\section{What the Second Wave Could Look Like}

If the second wave is to be more than critique, it could benefit from concrete starting points. Textual literacy has spent decades developing alternatives to measurement-centered approaches, and these offer visualization research somewhere to begin. We outline three, not as a complete program but as starting points for further work.

\subsection{Formative assessment}

In educational practice, formative assessment interprets evidence of learning to guide the next instructional steps; it does not record achievement or rank learners~\cite{Black:2009:DTA,Sadler:1989:FAD}. As Sadler puts it, it concerns \textit{``how judgments about the quality of student responses (performances, pieces, or works) can be used to shape and improve competence,''} unlike summative assessment, which reports a learner's achievement status~\cite{Sadler:1989:FAD}. It produces an evolving picture of what a learner can do across situations, not a single score. It also shifts where responsibility sits. Where the inherited frame asks how literate the individual is and treats any shortfall as their recorded property~\cite{Foucault:1977:DP}, formative assessment asks what should change around the reader: the visualization, the instruction, the pedagogy, or some combination. This moves the unit of improvement from the ranked person to the learning environment, loosening the individualizing logic that accompanied literacy's construction as a measurable property.

For visualization literacy, this suggests studying literacy in use, beyond what scores alone show. Imagine a student who pauses over whether a steep line reflects a real change or simply the axis they chose. A formative study would record that moment: what the student puzzles over, what they revise, and what they need from the visualization. Beasley et al. \cite{Beasley:2020:LPF} offer one instance: a visualization course built around feedback-guided continual improvement, in which students revise and reuse their work across projects in response to peer review, instead of being judged on a single submission. Such a study could draw on the artifacts this process already produces: draft visualizations, peer-review comments, and student reflections, used to guide instruction, not reduced to a score.

\subsection{Situated materials}

Critical literacy traditions argue that literacy materials and activities should be situated: grounded in the competencies people actually need in their own environments, cultures, and knowledge. They should not be assumed to generalize across contexts~\cite{Freire:2018:PO,Cazden:1996:PML}. As Street argues, literacy is best understood as \textit{``an ideological practice, implicated in power relations and embedded in specific cultural meanings and practices''} \cite{Street:1995:SLL}. Current visualization instruments often rely on Western cultural references and conventions, which can misjudge readers in other educational and cultural settings \cite{rakotondravony}. A VLAT item discussed by Nobre et al. \cite{nobre}, for instance, refers to the price of a barrel of oil; a test taker unfamiliar with that reference may be slowed or misled by it, so the item ends up measuring cultural knowledge alongside the ability to read the chart. Rakotondravony et al. \cite{rakotondravony} make the broader case that visualization practices are not neutral across language and culture. Translating a visualization's text into local languages, they note, \textit{``fundamentally ignore[s] the broader issue at hand — that monolingual research practices hide how the lived experiences of viewers may influence the ways in which they interact with visualizations,''} since it leaves the underlying conventions and assumptions untouched. The point is not that instruments should be translated better; no instrument is neutral with respect to context, and literacy measured in one setting may not mean the same thing in another. A second-wave agenda could build on this work, asking what visualization literacy looks like when grounded in the data practices and cultural settings of the communities involved, instead of measured against a default that was never neutral.

\subsection{Critical reading practices}

Critical pedagogy has developed practices that train learners to question what a text assumes, whose interests it serves, and what it leaves out \cite{Freire:2018:PO}. Applied to visualization, this means teaching people to read a chart and to interrogate it: why this chart and not another, what data it includes and excludes, what design choices direct attention, and whose questions it is built to answer. Some recent visualization research moves in this direction \cite{Hedayati:2024:FPP}; framing it explicitly as critical literacy work, with the pedagogical tradition behind it, would allow the field to build on decades of practical experience instead of starting over.

Formative assessment, situated materials, and critical reading practices are starting points, not a complete program. All three share a common orientation: each treats reading a visualization as something people do within a context and for a purpose, and asks what that activity needs. None fixes a person's ability as a number on a standardized task.

\section{Conclusion}

We have argued that visualization literacy is not a neutral measure of competence. It carries a history of quantification, individualization, and binary classification that the visualization field inherited with the term, and that history can determine what our assessments can and cannot encompass. What the historical lens makes visible is that literacy, as our field treats it, reduces a wide range of practices and competencies to a measured property and that the surrounding vocabulary then divides that property between those who have it and those who do not. The reduction was useful for nineteenth-century governance, and it has produced a careful first wave of visualization literacy research. But it is one historical operationalization of literacy: inherited, not chosen.

Naming the inheritance changes what is at stake. Refining a test is no longer automatic progress. The limits of an instrument become visible, including where a test built in one cultural setting misjudges readers in another. And a set of questions a score cannot answer comes back into question. Recent work has begun to ask these questions from within the field, Akbaba \cite{akbaba2026value} most directly; what this paper adds is the history that explains why measurement took hold and the tradition that offers an alternative. Recognizing the space the reduction leaves out opens ways forward: situated accounts of literacy as practice, engagement with the textual scholarship that has spent decades on this problem, and research programs that take measurement seriously without letting it stand in for literacy itself. None of this asks the field to give up measurement. It asks the field to hold measurement and meaning in the same frame, treating visualization literacy as a situated and contested practice, not a score, through which people make sense of a world increasingly run on data. That is what lets the field move past asking who has literacy and who lacks it, toward asking what we want literacy, in visualization, to be for.

\acknowledgments{
We would like to thank the Visualization and Interaction Design Lab at Linköping University for their valuable feedback on the ideas developed in this paper. We also thank the anonymous reviewers for their encouraging words and thoughtful engagement with our work. This research was funded by the Swedish Research Council (Dnr 2023-00933).}

\bibliographystyle{abbrv-doi}

\bibliography{template}
\end{document}